\documentclass[useAMS,usenatbib,psfig]{mn2e}
\input{psfig}

\title{Dissecting the star-formation history of starburst galaxies:
the case of NGC~7673}

\author[A. Pasquali and P. Castangia]{A. Pasquali$^{1}$\thanks{E-mail:
pasquali@mpia.de (AP); pcastang@ca.astro.it (PC)} and P. Castangia$^{2}$\\
$^{1}$ Max-Planck Institut f\"ur Astronomie, K\"onigstuhl 17,
D-69117 Heidelberg, Germany\\
$^{2}$ INAF - Osservatorio Astronomico di Cagliari, Poggio dei Pini,
Strada 54, 09012 Capoterra (CA), Italy}

\begin{document}

\date{}

\pagerange{\pageref{firstpage}--\pageref{lastpage}} \pubyear{2007}

\maketitle

\label{firstpage}

\begin{abstract}
We have collected archival data on NGC~7673 to constrain the 
star-formation history that produced the young star clusters and the field stellar
population in this galaxy during the last 2 Gyr. We have considered the
sample of 50 star clusters detected by HST/WFPC2 in the UV, V and I bands and
estimated their age, intrinsic reddening, and mass via comparison of their
colours with STARBURST99 models. We have found two prominent epochs of cluster
formation occurred about 20 Myr and 2 Myr ago, with somewhat minor events between
3 Myr and 6 Myr ago. The star clusters are characterised by an intrinsic 
reddening E(B-V) $<$ 0.4 mag and a mass lower than 2$\times$10$^6$ M$_{\odot}$.
Out of the 50 star clusters, we have selected 31 located within the boundaries
of the IUE large slit that was employed to obtain the spectrum of NGC~7673 between
1150 \AA\/ and 3350 \AA. For each cluster, we have built a synthetic spectrum 
corresponding to the age, mass and intrinsic reddening derived from the cluster 
colours, properly redshifted to NGC~7673. The spectra have then been
added together in a final, clusters integrated spectrum. This and the IUE and FUSE
spectra of NGC~7673 have allowed us to describe the star-formation history
of the unresolved stars in the field as either exponentially decaying or
multi-burst. In the first case, we have derived an e-folding time of 700 (900) Myr and
an initial star-formation rate of 16 (13) M$_{\odot}$~yr$^{-1}$ when 
the Fitzpatrick's (Calzetti's) extinction law is used. In the case of a 
multi-burst star-formation history, we have assumed 
that the same bursts which produced the star clusters 
built up also the field population. In this way, the field population turns out to
be composed by a young ($<$ 40 Myr) component 3 (2) times brighter than the star
clusters, and a component as old as $\sim$ 850 (450) Myr, about 200 (100) times more massive 
than the star clusters together. These star-formation histories fit equally well the 
observed UV spectrum of NGC~7673.  
\end{abstract}


\section{Introduction}

Star formation is one of the major drivers of galaxy evolution. As
already recognised by Hubble (1926), it is manifested in a variety of histories
and rates along the Hubble sequence, and in cosmic lookback time (Madau et al. 1996). 
In the present-day Universe, the star-formation
rate (SFR) spans about six orders of magnitude (in absolute terms, or about
2 per galaxy luminosity), from virtually zero in 
early-type galaxies up to $\sim$20 M$_{\odot}$~yr$^{-1}$ in gas-rich spirals, 
$\sim$100 M$_{\odot}$~yr$^{-1}$ in optically selected starburst galaxies, and 
$\sim$1000 M$_{\odot}$~yr$^{-1}$ in luminous IR starburst galaxies (cf. 
Kennicutt 1998). Locally, rapid star formation seems to occurr preferentially
in low-luminosity galaxies (cf. Cowie et al. 1996). These trends clearly
indicate that star formation is regulated by the galaxy gas content and
dynamical structure. The star-formation rate is observed to typically
increase with the amount of available gas and to be rather insensitive to
the dynamics due to a bar or a spiral arm structure. The latter mainly
affects the surface density of the star formation activity (cf. Knapen et al.
1992). On the contrary, tidal interactions and mergers between gas-rich systems
appear to enhance the star-formation rate; galaxies undergoing a relatively 
intense star-formation activity are indeed found to be morphologically disturbed 
(cf. Kennicutt 1998 
and references therein, Sanders \& Mirabel 1996). Gas loss due to ram-pressure
stripping can reduce this effect for cluster galaxies. In recent 
years, it has been claimed that AGN feedback (in its radio mode) can heavily
affect the star-formation activity of a galaxy to the extent of suppressing 
it (Croton et al. 2006). This would be the most effective way of truncating
star formation in early-type galaxies, in addition to gas exhaustion due to the
star-formation activity itself. Therefore, a complete description of galaxy 
evolution requires a detailed understanding of what can trigger and what can
repress the star-formation activity, and how star formation
takes place. It is well known, for example, that star formation changes in time 
with a different
pace in galaxies of different Hubble type: it runs out in early-type galaxies 
after 1 - 2 Gyr from the initial burst (Jimenez et al. 2006), while it is roughly 
constant or smoothly decaying in spirals (Searle et al. 1978, Kennicutt 1998, 
Kauffmann et al. 2003). A constant star-formation activity
has been also observed in a sample of starburst galaxies studied in ultraviolet
by Leitherer et al. (2002). All these star-formation histories (SFHs) are derived 
from the global photometric
and spectroscopic properties of galaxies, and their parameters (such as the age 
and duration of the burst, the star-formation rate) may change when some of 
these galaxies are analysed in greater detail. 

During the last decade, the {\it Hubble Space Telescope} (HST) has revealed
that a significant fraction of all stars produced during a burst
of star formation are formed within clusters. For example, Meurer et al. (1995)
found that the fraction of ultraviolet (UV) flux from star clusters varies between 
5\% and 52\% of the total galaxy UV emission, with an average of 20\%.
These star clusters are found to be blue and luminous, hence young (from few
10$^6$ to few 10$^8$ years) and with a mass typically between 10$^4$ and
10$^6$ M$_{\odot}$. For these reasons, they are commonly labelled as {\it young
massive clusters} (YMCs). YMCs are mostly detected in starburst galaxies, both major
and minor mergers, and, in a smaller extent, in spirals and dwarf-irregulars 
(Larsen 2004). Given  the significant contribution of YMCs to the 
stellar population of their host, their formation  comes to play
an important role in the global star-formation history of their parent galaxy.   
We could think of identifying, for example, two major components in the SFH of a 
starburst galaxy: a star-formation activity responsible for the unresolved stars
in the field and another one which built up the system of YMCs.   
We could then ask if and how these components correlate with each other: did
the field and clusters populations form in the same way, following the same
star-formation history? How intense was the clusters SFH with respect to the one
in the field? What are the typical timescales of the field and clusters SFHs? 
Would star clusters be able, upon disruption, to build up the field stellar
population? 
\par\noindent
The first studies by Boutloukos \& Lamers (2003) and de Grijs et al. (2003)
indicate that the cluster disruption timescale tends to be shorter (30 - 40
Myr) in interacting systems such as M~51 and M~82 and to increase up to few
Gyr in dwarf galaxies (8 Gyr in the SMC). According to these results, clusters
would then mostly  contribute to form the field population of interacting,
starburst galaxies. Chandar et al. (2005) estimate even shorter disruption
timescales, of the order of 10 Myr. 

Nowadays, the wealth of data in archives enables us to analyse in detail the SFH 
of a galaxy, and to address the questions above. The combination of
ground/space spectroscopy with the high, angular resolution imaging performed
by HST allows us to distinguish different components in the SFH of a galaxy,
and to characterise them separately with the support of state-of-the-art, theoretical 
models of stellar evolution. When applied to star-forming galaxies in different
dynamical state (non interacting, interacting, major and minor mergers), this
approach makes it possible to study the modes of star formation, their amplitude
and timescale, their dependence on galaxy mass and interactions, and the
subsequent evolution of a galaxy. 

In this paper, we make use of IUE, FUSE and HST UV spectra and 
photometry to ``dissect'' the star-formation activity occurred in the starburst 
galaxy NGC~7673 in the last 2 Gyr,  along the lines described above. 
NGC~7673 belongs to the class of the Luminous Blue Compact Galaxies characterised
by small sizes,  high surface brightnesses, strong emission lines and
blue colours (Gallagher et al. 1989). It exhibits a rather irregular and clumpy
morphology, but it also displays a rather regular, face-on disk kinematics 
(Homeier \& Gallagher 1999). It is in pair with NGC~7677 and no strong evidence
for an ongoing interaction has been found. Nevertheless, its similarity with
NGC~3310 (classified as a minor merger) in terms of optical and HI morphology
suggests that also NGC~7673 experienced a minor merger that probably triggered 
the starburst phase observed at the present time (Homeier \& Gallagher 1999).  

The paper develops as follow: data and models used in the analysis are presented 
in Sect. 2. The properties of the star clusters detected in NGC~7673 are 
described in Sect. 3, while the parameterization adopted to study the 
star formation in the field in Sect. 4. The results are then discussed in
Sect. 5, and the conclusions are drawn in Sect. 6. 
   
\section{The ingredients}
\subsection{UV spectroscopy}
NGC~7673 was first observed with the International Ultraviolet Explorer
(IUE) satellite in 1979 (programme JH174, PI: Heidmann) through the
SWP Camera. This programme made use of the large aperture,
characterised by an oval shape 10$'' \times$ 23$''$ in size, and acquired
a low-resolution (FWHM $\simeq$ 6 \AA) spectrum of the galaxy in the
range 1150 \AA\/ - 2000 \AA\/ with an exposure time of 18600 s (Benvenuti
et al. 1982). IUE visited NGC~7673 again, 
in 1981 (programme JH325, PI: Heidmann), through the LWP Camera,
which took a low-resolution (FWHM $\simeq$ 6 \AA) spectrum in the interval
1850 \AA\/ - 3350 \AA, using the large aperture and an exposure time of 24480 
s. Nineteen years later, the Far Ultraviolet Spectroscopic Explorer (FUSE) 
satellite acquired the spectrum of NGC~7673 between 900 \AA\/ and 1200
\AA\/ with the large, square aperture (30$'' \times$ 30$''$) and an
exposure time of 10038 s (programme A023, PI: Buat, Buat et al. 2002). 
\par
Benvenuti et al. (1982) identified absorption lines typical of OB stars
(e.g. SiIII $\lambda$1300, SiIV $\lambda$1400 and CIV $\lambda$1550) in the IUE 
spectrum of NGC~7673. The authors determined an
extinction of 0.31 mag in the B band and a dereddened luminosity
at $\lambda$ = 1550 \AA\/ of 1.9$\times$10$^{27}$ erg~s$^{-1}$~Hz$^{-1}$,
a factor of $\sim$100 larger than what measured for the HII region 30 Doradus
in the Large Magellanic Cloud (LMC). In addition, Buat et al. (2002) detected 
the presence of the OIV $\lambda$1032 line (due to massive O stars) and a 
decrease of the flux shortward of 1050 \AA. They interpreted these features
as the convolution of a multi-burst star-formation history (SFH, in the range
between 1 and 10 Myr) with the intervention of dust extinction. The authors 
also derived an extinction law almost constant between 1165 \AA\/ and 1040 \AA\/ 
and rising at shorter wavelengths.  
\par
We matched the IUE and FUSE spectra over the range 1150 \AA\/ -
1200 \AA\/ by scaling the latter spectrum by a factor of 0.5 to compensate
for the larger aperture used by FUSE. 

\subsection{UV imaging}
Between 1996 and 1997, the Hubble Space Telescope (HST) imaged NGC~7673
with the Wide Field and Planetary Camera 2 (WFPC2) in the visible with
the F555W and F814W filters, and in the UV through the F255W filter
(programme ID: 6870, PI: Trauger). The observations were carried out with
exposure times varying from 2300 s in
the UV to 800 s~for the F555W filter and 1200 s for F814W. These images
were analysed by Homeier et al. (2002), who identified 268 young star clusters,
mostly located in clumps. The F255W image of NGC~7673 (from Homeier et al.)
is shown in Figure 1 with the IUE large aperture overlaid and the star 
clusters labelled.

\begin{figure*}
\centerline{\psfig{figure=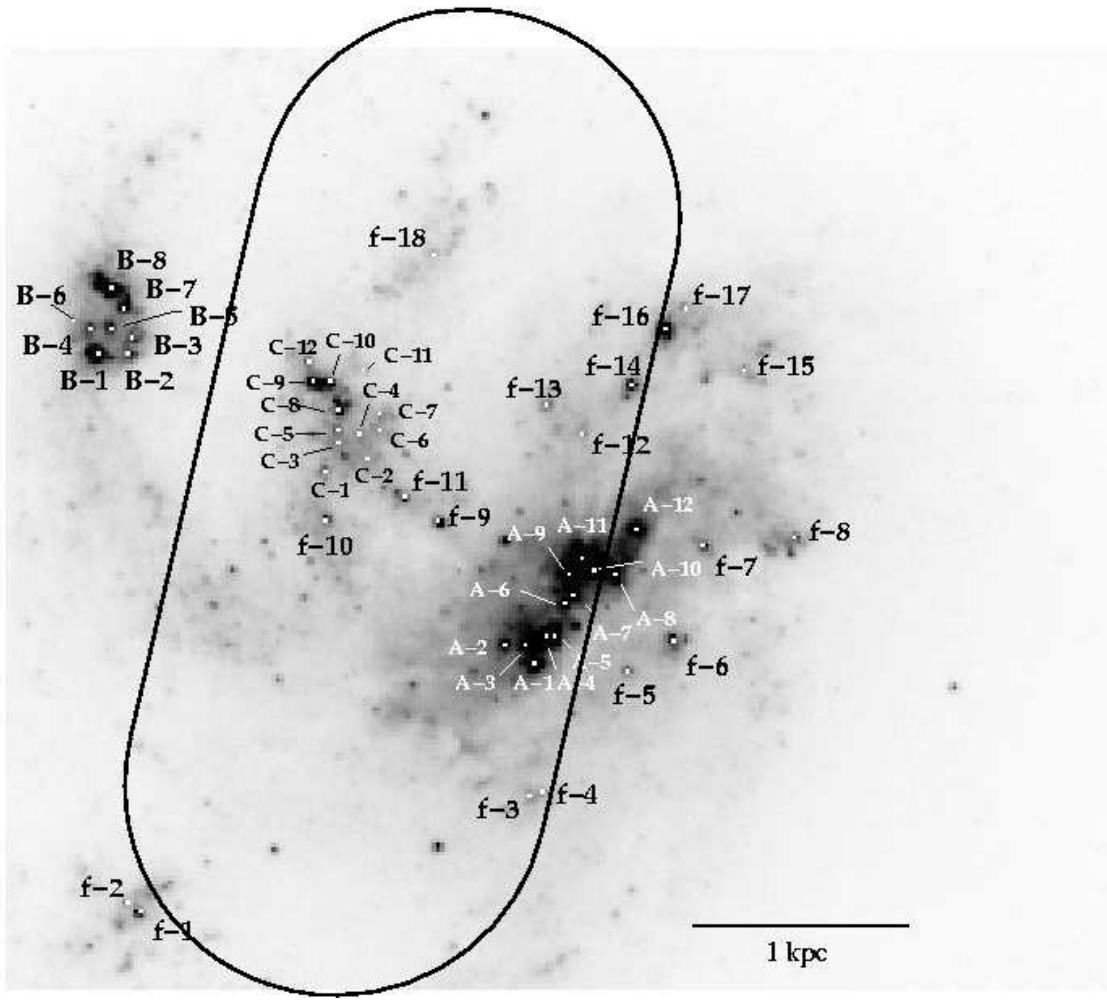,width=15cm}}
\caption{The WFPC2/F255W image of NGC~7673 with superposed the IUE large
aperture. Clumps and individual star clusters are 
labelled according to Homeier et al. (2002). North is up and East to
the left}
\end{figure*}

Fifty clusters have been detected in all the three filters, and their colours 
allowed Homeier et al. to estimate
an age between 5 and 20 Myr. In particular, younger clusters seem to be
located in clumps A, B, C, D while older clusters are preferentially 
associated with clump A. The authors measured a clusters contribution to
the total light of a clump between 16$\%$ and 33$\%$.

\subsection{The models}
In order to determine the star-formation history of the star clusters 
and the stars in the field of NGC~7673, we have to compare their
observed photometry and spectra with synthetic stellar populations
properly convolved with a specific reddening law. We ran STARBURST99
(Leitherer et al. 1999) to compute the spectrum of an instantaneous
starburst (i.e. a single stellar population) with a total mass of
10$^6$ M$_{\odot}$ as a function of its 
age up to 2 Gyr, when its UV emissivity becomes negligible. For this purpose, 
we assumed the default parameters of STARBURST99 except for the metallicity
that we chose to be LMC-like. This value was estimated by Duflot-Augarde \& 
Alloin (1982) from the oxygen abundance of the interstellar medium. We adopted 
a multi-power law
for the initial mass function (IMF) in order to approximate a Kroupa (2001)
IMF; an exponent of 1.3 was chosen for stars with a mass in the range  0.1 
M$_{\odot}$ - 0.5 M$_{\odot}$, and 2.3 for the interval 0.5 M$_{\odot}$ -
100 M$_{\odot}$. Only stars with a mass in the interval 8 M$_{\odot}$ to
120 M$_{\odot}$ were selected to produce supernovae and black holes. We 
chose the stellar wind and mass loss to be computed with the theoretical model,
which solves for the radiative transfer and momentum equation in the 
stellar wind (cf. Leitherer et al. 1992).
With the same parameters described above, we also ran STARBUST99 in order
to synthesize the spectrum of a stellar population continuously forming 
stars (with a constant star-formation rate of 1 M$_{\odot}$~yr$^{-1}$) as
a function of age up to 2 Gyr.  

In order to assess the dependence of our results on the evolutionary
tracks used in STARBURST99, we computed synthetic spectra from the
Geneva standard tracks (GST), the Geneva tracks with high mass-loss (GHT)
and the Padova tracks including the AGB evolutionary phase (PAT). The
Geneva tracks are fine-tuned on the evolution of massive stars (M$_* >$
10 M$_{\odot}$) and on the treatment of their  winds and mass loss
(cf. Meynet et al. 1994), while the Padova models treat more extensively
the evolution of low- and intermediate-mass stars through their post 
main-sequence phases (cf. Bertelli et al. 1994). 

In the case of an istantaneous starburst, we redshifted the spectra synthesized
with the GST, GHT and PAT evolutionary tracks to NGC~7673 ($z$ = 0.011) and
convolved them by the total throughputs of the F255W, F555W and F814W filters,
in order to obtain synthetic colours to compare with the observed ones as a
function of age.

\section{The star clusters properties}
Out of the catalogue published by Homeier et al.,
50 star clusters have both the (F255W-F555W) and the (F555W-F814W)
colours measured. Thirty-one of these clusters are placed within the IUE large
aperture.  From these two colours, we estimated the star clusters properties, 
i.e. age, mass, and intrinsic reddening, using the technique described in 
Pasquali et al. (2003). Briefly, we fitted the observed colours, corrected 
for the Galactic E(B-V)$_G$ = 0.043 (Schlegel et al. 1998) with the Cardelli
et al.'s (1989) extinction law, to the synthetic colours of a single stellar
population with a total mass of 10$^6$ M$_{\odot}$ and a LMC-like metallicity.
Except for stochastic effects associated with small numbers of very luminous
stars, synthetic colours are not mass sensitive, but mainly depend on age
and intrinsic reddening within NGC~7673. This reddening is left as a free
parameter, varying between E(B-V)$_i$ = 0 and 5 mag, with a step of 0.1 mag.  
For a given E(B-V)$_i$, the synthetic colours at each time point are reddened 
with a specific extinction law and a $\chi^2$ is calculated as the sum of
the differences between the observed and the reddened synthetic colours
normalized by the observational errors. The minimum value of $\chi^2$,
$\chi^2_{min}$, defines
the best-fitting age, intrinsic reddening, and synthetic V magnitude. 
Following the correction for the best-fitting intrinsic reddening, the
difference between the observed and synthetic V magnitudes is used to
estimate the cluster mass via the scaling relation:
\[
M_{cl} = 10^{-0.4(V_{obs} - V_{syn})} \times 10^6~M_{\odot}.
\]
The errors on each best-fitting parameter are given by the minimum 
and maximum value among all the fits  which realise  $\chi^2 <$
1.5$\chi^2_{min}$ (corresponding to about 2$\sigma$). 
\par
We have modelled the intrinsic reddening of NGC~7673 with three different 
extinction laws: the one proposed by Calzetti (2001) for starburst
galaxies, the extinction law derived by Cardelli et al. for the Milky Way and
that computed for the LMC by Fitzpatrick (1986). The
Fitzpatrick's law has been extraploted at $\lambda <$ 1000 \AA\/, while
the Cardelli et al.'s and Calzetti's curves have been extended shortward of 
1200 \AA\/ with the Sasseen et al.'s (2002) and Leitherer et al.'s (2002) 
extinction laws, respectively. Moreover, the Fitzpatrick's curve has been extended at 
$\lambda >$ 3000 \AA\/ with the Cardelli et al.'s law; both curves assume
$\frac{A_V}{E(B-V)}$ = 3.1. 
\par
At the end, the fitting technique described above produces for each star 
cluster a 3 $\times$ 3 grid of results (for 3 sets of evolutionary tracks 
and 3 extinction laws), which allow us to discuss any systematic effect 
introduced by different evolutionary tracks, each of them reddened with 3
different extinction laws.

\subsection{Ages, intrinsic reddening, and masses}
The histogram of the clusters ages is shown in Figure 2, for each 
combination of a set of evolutionary tracks with an assumed extinction law.
In all cases, the grey shaded histogram represents the 50 star clusters
detected in the 3 WFPC2 filters, while the one shaded in black outlines
the age distribution of the 31 star clusters included in the IUE 
aperture. As expected, the cluster age largely depends on the evolutionary
tracks chosen for STARBURST99, and is far less sensitive to the adopted
extinction law. Irrespective of how reddening is treated, the GST and
GHT evolutionary tracks clearly indicate two epochs of cluster formation,
about 20 and 2 Myr ago, with minor events in between, about
3 - 6 Myr ago. On the contrary, the PAT evolutionary tracks point to a
somewhat more continuous cluster formation, which took place between
2 and 100 Myr ago, with  two major bursts 4 and 10 Myr ago.  


Star clusters as young as 10$^4$ yr could be unresolved HII regions. 
The average error on these ages is about 8 Myr. 
As discussed in Pasquali et al. (2003, 2004), the accuracy of the
age determination improves significantly when three or more colours are used
in the analysis, and cover a large baseline extending from ``age-sensitive''
(UV) to ``dust-sensitive'' (I,IR) wavelengths.
The individual age uncertainties are plotted in Figure 3, where the ages 
derived with the GST and PAT tracks are compared with the values obtained with 
the GHT models for each different extinction law. Clearly, the rather large
errors do not allow us to discriminate among the various evolutionary tracks, 
and make plausible both the cluster formation histories described above.  
\par\noindent
The intrinsic cluster reddening does not exceed the value E(B-V)$_i$ $\simeq$
0.4 mag (see Figure 4) and, for a large number of star clusters, is 0 $\leq$ 
E(B-V)$_i$ $\leq$ 0.25 mag, independently of the adopted extinction law. 
No significant dependence of the best-fitting reddenings on the evolutionary tracks is
observed. The average uncertainty on E(B-V)$_i$ is about 0.05 mag. 

\setcounter{figure}{7}
\begin{figure*}
\centerline{\psfig{figure=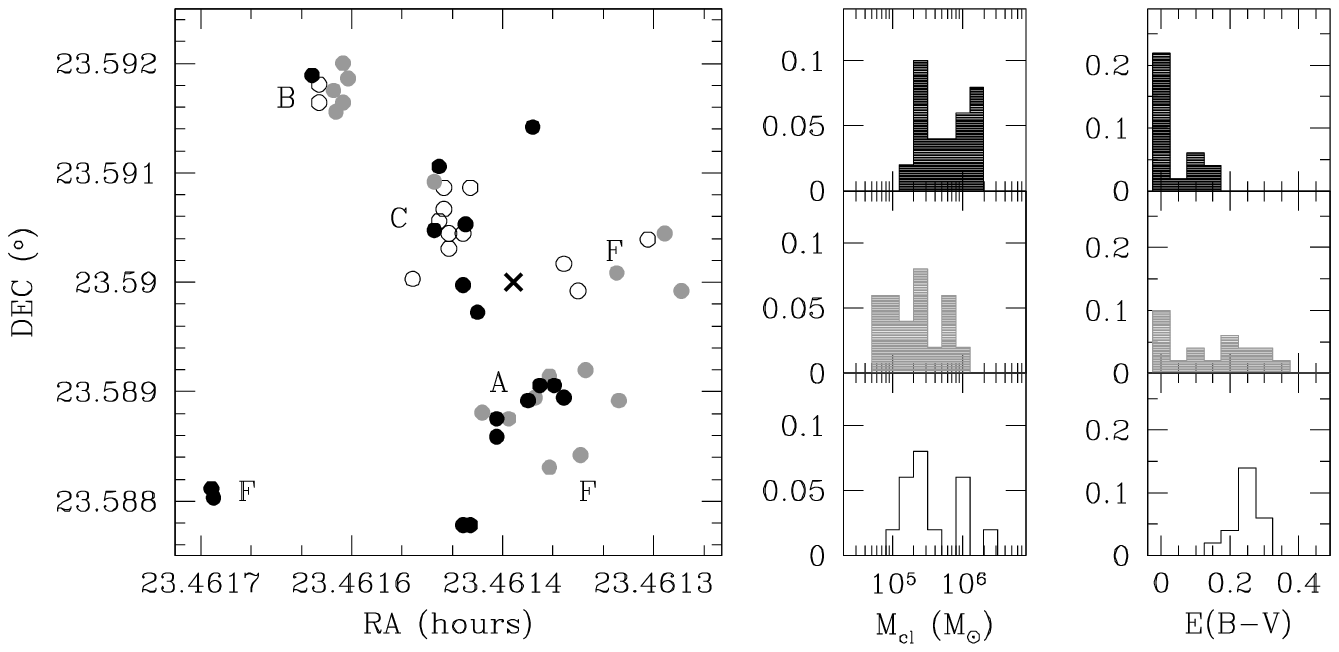,width=15cm}}
\caption{The spatial distribution of the star clusters as a function of
their age. Open circles represent clusters younger than 2.5 Myr, the grey
circles stand for intermediate age clusters (between 2.5 and 8 Myr), while
the black circles indicates the star clusters older than 8 Myr.
The distributions in mass and reddening of the three clusters age-groups
are plotted on the right-hand side of the Figure, and are normalized to
50, the total number of clusters detected in F225W, F555W and F814W.
Ages, masses and reddening were
computed with the GHT tracks and with the Fitzpatrick´s extinction law.
The clumps of star clusters are also labelled according to Homeier et al.
(2002). The cross indicates the galaxy centre.}
\end{figure*}

\par\noindent
Given that the clusters mass is computed from the observed absolute magnitude
corrected for the best-fitting reddening, it is itself sensitive to the
adopted extinction law and 
is generally lower than $\sim$2 $\times$ 10$^6$ M$_{\odot}$. 
The individual mass errors are shown in Figure 7 for the different evolutionary 
tracks and extinction laws; on average they are about 3$\times$10$^5$ M$_{\odot}$.
\par 
We have splitted the full sample of star clusters into three groups on the
basis of their estimated age as obtained from the GHT tracks and the 
Fitzpatrick's extinction law: the young sample with ages $\leq$ 2.5 Myr, the
intermediate one between 2.5 and 8 Myr, and the old sample with ages $>$ 8 
Myr. These groups are
represented in Figure 8 with open, grey and black circles, respectively,
and are plotted as a function of their RA and DEC (J2000) coordinates. The
clumps originally identified by Homeier et al. are also labelled with
the letters A, B and C, while F indicates the clusters in the field. Modulo
the large age uncertainty discussed above, the star clusters of NGC~7673 
seem to get younger as their displacement in the NE direction from 
clump A increases.
The distributions of these groups in mass and intrinsic reddening are also
traced in Figure 8. It could be observed that, for increasing age, the clusters
tend to be less obscured and more massive; but this trend can well be due 
to the depth of the data, since, for a given U magnitude,
intrinsically older clusters need to be more massive and less extinguished than
the younger ones in order to be detected. These results also hold for the
Cardelli et al.'s and Calzetti's laws.

\subsection{UV spectra}
The age and mass derived from the photometry can now be
used to construct, with the help of STARBURST99, the UV spectrum of each
cluster for different evolutionary tracks and extinction laws. The correction
for dust obscuration is twofold: first, each spectrum at restframe wavelength
is reddened by E(B-V)$_i$ via a given extinction law assumed to represent the 
obscuration within NGC~7673. The spectrum is then redshifted to the
distance of NGC~7673 ($z$ = 0.011) and estinguished by the Galactic E(B-V)$_G$ 
through the 
Cardelli et al.'s extinction law. Clearly, this procedure can be used with
the best-fitting values of age, mass and intrinsic reddening as well as
with the lower and upper limits computed for these parameters. 
The spectra of all clusters are then
summed together in what we refer to as {\it the clusters integrated spectrum}
and compared with the observed FUSE/IUE spectrum of NGC~7673, in order to
check the contribution of the clusters to the total emission of the galaxy 
and the systematic effects due to the observed photometry and the fitting
technique. 

The clusters integrated spectrum computed from the GHT tracks and for
different extinction laws is plotted in Figure 9. Here, the black solid
line represents the clusters integrated spectrum obtained for the
best-fitting values of age, mass and E(B-V)$_i$, while the black short-
and long-dashed lines trace, respectively, the clusters integrated spectrum 
derived for the lower and upper limits of these parameters. 
The observed IUE spectrum of the galaxy is shown with a grey solid 
line.  The clusters integrated spectrum is on average 2 mag fainter than
the observed IUE spectrum of the galaxy; the inclusion of errors
in the clusters intrisic reddening or mass is able to make it about 1
mag fainter than the galaxy spectrum (cf. lower, right-hand side
panel of Figure 9). At the mean wavelength of the
WFPC2 F255W filter (2578 \AA), the clusters integrated spectrum contributes 
about 20$\%$ to the galaxy light (between 15$\%$ and 38$\%$, when errors 
are taken into account) independently of 
any of the extinction laws in use. This contribution turns out to be 
insensitive also to the adopted set of evolutionary tracks. 
Using the F255W image of NGC~7673, Homeier et al. 
estimated a mean cluster contribution of 24$\%$, in agreement with what
we measured for the portion of galaxy included in the IUE slit. 
We then can not use this value to constrain the extinction law that
best represents the dust obscuration within NGC~7673 among the three
used in this paper.  Since Duflot-Augarde \& Alloin (1982) and
Gallagher et al. (1989) measured for NGC~7673 a LMC-like metallicity,
we only discard the extinction law by Cardelli et al., and from now on we
will apply the Fitzpatrick's and Calzetti's laws. 

\section{Star formation in the field}
As Figure 9 indicates in its left-hand side panels, star clusters
in NGC~7673 contribute only $\sim$ 20$\%$ to the UV emission of their parent 
galaxy. Homeier et al. determined the completeness of their clusters sample
in the F555W filter as a function of magnitude and background level.
The latter is indeed quite variable given the clumpy morphology of NGC~7673. 
For the highest background level, the authors estimated a recovery fraction
of about 50$\%$ at $m_{F555W}$ = 26 mag, increasing to 98$\%$ at $m_{F555W}$ 
= 21.5 mag. Since the clusters involved in our analysis (those with two 
measured colours) are brighter than $m_{F555W} \simeq$ 22 mag, we assume
for them a completeness level larger than 90$\%$. This should guarantee
a high statistical significance to any further analysis of the clusters 
properties.
\par
The next step is then to constrain the nature of the remaining 80$\%$ of
the flux in the FUSE/IUE spectrum of NGC~7673, which we assume to be arising 
from the stellar population in the field. Specifically, we aim at deriving 
the star-formation history 
(SFH) of the field, and checking whether the field is consistent with being 
composed by stars 
born in clusters later disrupted by their gravitational interaction with 
the galaxy. We have proceeded by assuming a specific SFH for the field,
which has been {\it i)} reddened according to the prescription in Sect. 3.2
and with the Fitzpatrick's and Calzetti's laws as representing the intrinsic dust absorption
of the galaxy; {\it ii)} summed to the clusters integrated spectrum 
(computed with the Fitzpatrick's and Calzetti's laws) and 
{\it iii)} fitted to the observed FUSE/IUE spectrum of the galaxy with a 
minimum $\chi^2$ approach. As discussed by Meynet et al. (1994), the 
Geneva tracks with a high mass loss rate are preferable in modelling the
evolution of massive stars (i.e. young ages of a stellar population), since
they better reproduce the low luminosity observed for some Wolf-Rayet stars,
the surface chemical composition of WC and WO stars, and the ratio of blue
to red supergiants in the star clusters in the Magellanic Clouds. For
this reason, the modelling of the field SFH in  
NGC~7673 has been restricted only to the GHT tracks. 

\subsection{Continuous star formation} 
For our first attempt, we have used the STARBURST99 spectra of a population
continuously forming stars with a constant rate of 1 M$_{\odot}$~yr$^{-1}$ 
computed as a function of age (see Sect. 2.3). The best
fit to the observed FUSE/IUE spectrum of NGC~7673 is shown in panels {\it a}
and {\it d} of Figure 10, for the Fitzpatrick's and Calzetti's extinction
law respectively. Here, the dashed grey line traces the clusters integrated spectrum 
(labelled with C), the 
dot-dashed grey line represents the field spectrum resulting from
its continuous SFH (labelled with F), the solid grey line is the sum of the clusters
integrated spectrum with the field best-fitting SFH. The observed spectrum
is shown in black; the vertical errorbars
indicate a 1$\sigma$ error on the measured FUSE/IUE fluxes. 
The field best-fitting SFHs are described by an age of about 650 Myr (Fitzpatrick)
or 610 Myr (Calzetti), an intrinsic reddening E(B-V)$_i$ $\simeq$ 0.5 mag and a 
star-formation rate of about 20 M$_{\odot}$~yr$^{-1}$ (Fitzpatrick) or 
26 M$_{\odot}$~yr$^{-1}$ (Calzetti). The reduced $\chi^2$ realized by these fits is
2.5 and 0.8 for the Fitzpatrick's and Calzetti's extinction law respectively. 
In the case of the Fitzpatrick's law, the best
fit underestimates the flux shortward of 1300 \AA\/ and overestimates the bump at 2200
\AA; on the contrary, the fit obtained with the Calzetti's law nicely reproduces
the observed FUSE/IUE spectrum of the galaxy. Nevertheless, both extinction laws 
give a reddening for the field larger than the range computed for the star clusters, and 
a star-formation rate dramatically higher than the value derived from the H$\alpha$
emission line ($\simeq$ 1 M$_{\odot}$~yr$^{-1}$, Schmitt et al. 2006).

\setcounter{figure}{9}
\begin{figure*}
\centerline{\psfig{figure=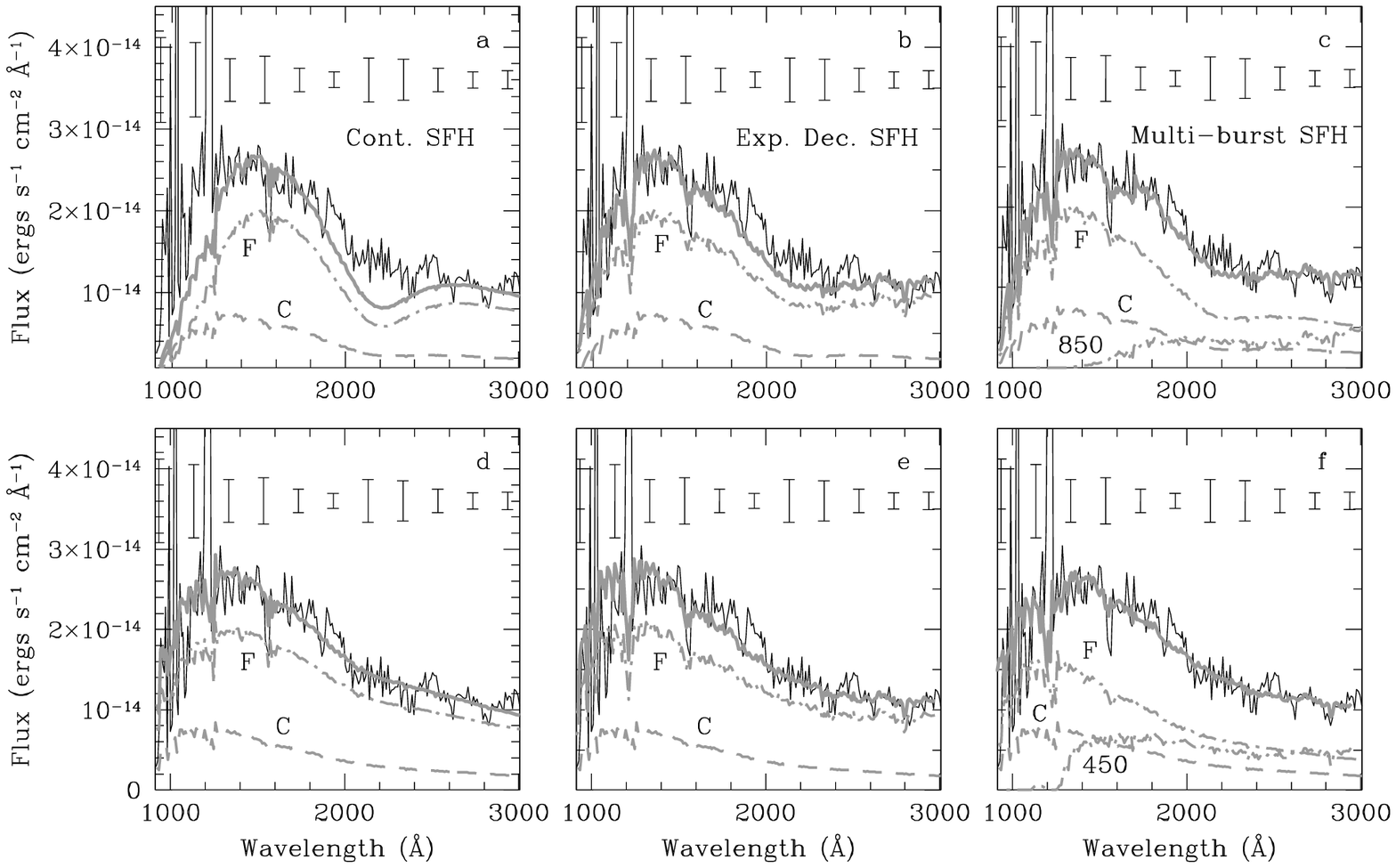,width=15cm}}
\caption{The observed FUSE/IUE spectrum of NGC~7673 (solid black line) 
compared with the best-fitting spectral energy distribution (SED, solid grey line),
that is the sum of the clusters integrated spectrum with the field SED. 
The clusters integrated spectrum is shown with a dashed grey line, while 
the field SED with a dash-dotted grey line. The field SED
has been obtained under the assumption
of a continuous star formation in panels {\it a} and {\it d}, exponentially
decaying star formation in panels {\it b} and {\it e}, and multi-burst
SFH in panels {\it c} and {\it f}. 
The best-fitting SFHs in panels {\it a}, {\it b} and {\it c} 
are based on the Fitzpatrick's extinction law, while those in panels {\it d},
{\it e} and {\it f} have been computed with the Calzetti's law.}   
\end{figure*}

\subsection{Exponentially decaying star formation}
The STARBURST99 models of a single stellar population as a
function of age calculated for the star clusters in Sect. 3 can be 
used to define an exponentially decaying SFH for the field:
\[
SFR(t) = SFR_o ~ e^{\frac{-(t_o - t)}{\tau}}~ (M_{\odot}~yr^{-1}), 
\]
where $t$ is the lookback time from now at $t$=0 to the initial burst 
occurred $t_o$ = 2 Gyr ago with a star-formation rate $SFR_o$, 
$\tau$ is the e-folding time needed for
the SFR to decrease by a factor $e$, and $SFR(t)$ is the star-formation 
rate at lookback time $t$. The choice of $t_o$ = 2 Gyr is dictated
by the fact that the UV emissivity of a stellar population becomes negligible
at ages older than this value. Fitting the above formula to the observed
FUSE/IUE spectrum of NGC~7673 constrains the $e$-folding time of the 
star-formation activity in the field, the field reddening, the initial and 
present-day star-formation rate. The best-fitting spectrum of the field
is plotted in panels {\it b} and {\it e} of Figure 10 for each of the adopted
extinction laws. Both SFHs match the observed spectrum quite tightly
with a reduced $\chi^2$ of 1.6 and 1.2 for the Fitzpatrick's and Calzetti's
extinction laws respectively. Once again, the spectrum of the field
computed with the Fitzpatrick's law overestimates (at 1.5$\sigma$ level) 
the bump at 2200 \AA. The field stellar population is
best reproduced by a star-formation history which begun with a burst
about 2 Gyr ago and an initial rate of $\simeq$ 16 M$_{\odot}$~yr$^{-1}$
(Fitzpatrick) or 13 M$_{\odot}$~yr$^{-1}$ (Calzetti), it 
fades with an $e$-folding time of $\simeq$ 700 Myr (Fitzpatrick) or 900 Myr 
(Calzetti), so that the present-day star-formation rate is of $\simeq$ 1 
M$_{\odot}$~yr$^{-1}$. The dust obscuration in the field corresponds to 
E(B-V) $\simeq$ 0.13 mag, consistent with the average reddening determined for 
the star clusters.   

\subsection{Multi-burst star formation}
A ``multi-burst'' scenario is a third, alternative scenario for the SFH
of the field in NGC~7673. Here, the assumption is that the bursts
of star formation, which produced the observed star clusters, also built 
up the field stellar population. We have thus scaled
the clusters integrated spectrum to properly fit the FUSE/IUE spectrum of 
NGC~7673 at $\lambda <$ 1500 \AA\/ and added an older, single stellar
population with a total mass to match the observed spectrum at redder 
wavelengths. 
The best-fitting spectrum is shown in panels {\it c} and {\it f} of 
Figure 10. The best-fitting SFH requires the
field to have a young component (with an age $<$ 40 Myr)
$\simeq$ 3 (Fitzpatrick, or 2 with Calzetti's law) times brighter than the 
clusters (assuming the same reddening distribution), and an unobscured component as old 
as 850 Myr (Fitzpatrick) or 450 Myr (Calzetti). This old population turns out
to be as massive as 4$\times$10$^9$ M$_{\odot}$ (Fitzpatrick) or 2$\times$10$^9$ M$_{\odot}$ 
(Calzetti).
The reduced $\chi^2$ realized by this best-fitting spectrum is 1.2 (Fitzpatrick)
or 0.9 (Calzetti),
comparable with the value obtained for the exponentially decaying SFH.
In the case of the Fitzpatrick's extinction law, the multi-burst SFH fits the bump at 2200 
\AA\/ slightly better than the exponentially decaying SFH.

\subsection{Comparisons}
The best-fitting spectra obtained in Figure 10 clearly rely on the
accuracy of the clusters integrated spectrum, hence on the accuracy 
achieved in estimating the age, reddening and mass of the star clusters from 
their photometry. We have shown in Figure 9 how the clusters integrated 
spectrum varies as a function of the errors implied by our ``photometric''
clusters dating technique. We can thus repeat the above analysis of the field
SFH using the variations of the clusters integrated spectrum, in order to
assign an uncertainty to the field best-fitting parameters. For example,
in the case of a continuous SFH, the age of the field stellar population
turns out to be 650$_{-40}^{+550}$ Myr (Fitzpatrick) or 
610$_{-20}^{+490}$ Myr (Calzetti), and the star-formation rate
20$_{-1}^{+13}$ M$_{\odot}$~yr$^{-1}$ (Fitzpatrick) or
26$_{-4}^{+30}$ M$_{\odot}$~yr$^{-1}$ (Calzetti),
while the reddening does not vary
from E(B-V) = 0.5 mag. In the hypothesis of an exponentially decaying SFH,
we obtain $\tau$ = 700$_{-150}^{+190}$ Myr (Fitzpatrick) or
900$_{-100}^{+100}$ Myr (Calzetti), an initial star-formation rate
of 16$_{-3}^{+17}$ M$_{\odot}$~yr$^{-1}$ (Fitzpatrick) or
13$_{-3}^{+16}$ M$_{\odot}$~yr$^{-1}$ (Calzetti), and E(B-V) = 0.12$_{-0.04}^{+0.24}$
mag (Firzpatrick) or 0.13$_{-0.01}^{+0.15}$ mag (Calzetti). 
Finally, for the multi-burst/Fitzpatrick SFH, the field young stellar population ranges
from being a factor of 2 fainter than the star clusters to being about 4 times 
brighter, while the older component turns out to have a mass of 4$_{-1}^{+3} 
\times$ 10$^9$ M$_{\odot}$ and a reddening between 0 and 0.04 mag. In the
case of the multi-burst SFH computed with the Calzetti's law, the field 
can be as bright as the clusters or 3 times brigther. The mass of the older component
ranges between 1 and 2 $\times$ 10$^9$ M$_{\odot}$ and its reddening is negligible. 
We then conclude from these comparisons that, within the errors of our clusters dating
technique, the SFHs obtained with the Fitzpatrick's law are consistent with those
computed assuming the Calzetti's law. For this reason, in what follows we will make use
of quantities averaged between the estimates given by the two extinction laws.

\section{Discussion}
Inspection of Figure 10 immediately rules out that
the field of NGC~7673 is undergoing a continuous star-formation activity 
with a constant rate of about 20 M$_{\odot}$~yr$^{-1}$ or more. The other two
scenarios, of an exponentially decaying and a multi-burst SFH, are in
better agreement with the observed FUSE/IUE spectrum of NGC~7673 and its
H$\alpha$ SFR, and fit this spectrum equally well. Both parameterized SFHs can be used to
constrain the radiation field and the stellar mass of the galaxy, and to investigate 
the nature of the field as built up by disrupted clusters.

\subsection{Radiation field}
In the hypothesis of an
exponentially decaying SFH, the field stellar population would form stars
with a present-day rate of 1$_{-0.5}^{+2.5}$ M$_{\odot}$~yr$^{-1}$. 
A similar value has been computed by Schmitt et al. (2006) from the H$\alpha$
emission of the galaxy, a factor of about 4 lower than the star-formation 
rate derived from far-IR observations ($\simeq$ 4.9 M$_{\odot}$~yr$^{-1}$). 
It has to be specified that the
value derived by Schmitt et al. is a lower limit to the true star-formation 
rate in the optical, since the authors did not correct the observed H$\alpha$
flux for intrinsic reddening but only for the Galactic extinction. Since 
this H$\alpha$ flux was originally measured by McQuade et al. (1995) together
with the H$\beta$ flux, it is then possible to estimate the reddening
affecting the nebular emission lines. Under recombination case B, the
observed $\frac{H\alpha}{H\beta}$ ratio leads to a nebular reddening
$c$ = 0.78, which, following Kaler \& Lutz (1985), can be translated into
E(B-V) $\simeq$ 0.5 mag (0.54 mag including the Galactic extinction). 
Applying the Fitzpatrick's extinction law gives
a dereddened H$\alpha$ flux $\simeq$ 2$\times$10$^{-12}$ erg~cm$^{-1}$~s$^{-1}$,
hence, for a distance of 49 Mpc (Homeier et al. 2002), a luminosity
$L(H\alpha) \simeq$ 6.6$\times$10$^{41}$ erg~s$^{-1}$. Such a nebular
reddening seems to be quite high when compared with the values derived for 
the star clusters (see Figure 4) and the stellar continuum. A stellar E(B-V) =
0.12 mag as estimated for the exponentially decaying SFH (0.16 mag adding
the Galactic extinction) would translate into a reddening of $\sim$ 0.36 mag
for the gas, hence it would give a 
H$\alpha$ luminosity of $\simeq$4.2$\times$10$^{41}$ erg~s$^{-1}$.   
When using the formula:

\[
SFR = 7.9\times10^{-42}L(H\alpha)~M_{\odot}~yr^{-1}
\]
(cf. Kennicutt 1998), these two $L(H\alpha)$ values give a star-formation rate 
between 3.3 and 5.2 M$_{\odot}$~yr$^{-1}$, still
consistent with the results of our best-fitting, exponentially decaying
SFH.

A reddening-free H$\alpha$ emission provides us with an independent check 
of the reliability 
of our SED modelling. McQuade et al. (1995) measured the H$\alpha$ flux of
NGC~7673 within a circular aperture equivalent in area to the IUE slit. As
shown by Homeier and Gallagher (1999), the H$\alpha$ emission is spatially
associated with the clumps (A, B and C in Figure 1) detected at other 
wavelengths. Therefore, any difference in the positioning of the aperture 
with respect to the IUE slit becomes quite critical, and may affect the 
measured line flux. In what follows, we assume the H$\alpha$ flux 
of McQaude et al. to be representative of the emission occurring
within the IUE slit, and we use it to estimate the balance between the
ionizing photons emitted by the cluster and field stars and those needed
to sustain the H$\alpha$ luminosity. For a given IMF, 
the number N(H$^0$) of ionizing photons emitted per second at $\lambda <$ 
912 \AA\/ critically depends on the mass and age of the stars and less 
significantly on their metallicity (Leitherer \& Heckman 1995). Stars of
OB spectral types and Wolf-Rayet stars dominate N(H$^0$), which, in the
case of an instantaneous burst, peaks during the first 5 Myr and 
sharply declines at older ages. The best-fitting exponentially decaying SFHs,
produce N(H$^0$)$_{field}$ $\simeq$ 1.5$\times$10$^{53}$ photons per second. 
This has to be added to the 
contribution from the star clusters alone, N(H$^0$)$_{clusters}$ $\simeq$
3.4$\times$10$^{53}$, in order to derive a total N(H$^0$)$_{total}$ of
4.9$\times$10$^{53}$ photons per second. The uncertainties in our cluster
dating technique and SED modelling place a lower and upper limit to
N(H$^0$)$_{total}$ of 2.5$\times$10$^{53}$ and 1$\times$10$^{54}$ s$^{-1}$,
respectively. The number of ionizing photons
can also be computed from the measured H$\alpha$ luminosity via the
relation:  

\[
N(H^0) = 7.35\times10^{11}L(H\alpha)~s^{-1}
\]
(cf. Leitherer \& Heckman 1995). We derive N(H$^0$)$_{H\alpha}$ between
3.1$\times$10$^{53}$ [for $L(H\alpha)$ = 4.2$\times$10$^{41}$ erg~s$^{-1}$] 
and 4.9$\times$10$^{53}$ photons~s$^{-1}$ [for $L(H\alpha)$ = 
6.6$\times$10$^{41}$ erg~s$^{-1}$], consistent with N(H$^0$)$_{total}$.

In the case of a multi-burst SFH, the best-fitting SEDs in Figure 10 
give N(H$^0$)$_{multi-burst} \simeq$ 1.1$\times$10$^{54}$ photons~s$^{-1}$. 
Once all the uncertainties on the best fit are taken into account,
N(H$^0$)$_{multi-burst}$ turns out to vary between 2.9 $\times$10$^{53}$ and
2.5$\times$10$^{54}$ s$^{-1}$ similarly to both N(H$^0$)$_{total}$
and N(H$^0$)$_{H\alpha}$. Given that the multi-burst SFH is based on star
clusters that are younger than the exponentially decaying SFH, it thus 
predicts a higher number of ionizing photons. 

In conclusion, the stars produced in clusters and in
the field via either an exponentially decaying or a multi-burst SFH
emitt a number of ionizing photons large enough to sustain the H$\alpha$
luminosity. The errors in the modelling of the observed UV stellar continuum
would allow the ionizing photons to escape with an average fraction of 
about 50$\%$ (cf. Weedman 1991).

\subsection{Stellar mass}
The star clusters within the IUE slit lock up a total mass M$_{clusters}$ =
2$_{-0.3}^{+2.2}\times$10$^7$ M$_{\odot}$.
The stellar mass of the field
population clearly depends on the parameterization of its star-formation 
history. In the case of an exponentially decaying SFH, the stellar mass in
the field is M$_{field}^{ex}$ = 1.1$_{-0.2}^{+1.5}\times$10$^{10}$ M$_{\odot}$.
Since M$_{clusters}$ is about only 0.2$\%$ of M$_{field}^{ex}$, the total 
stellar mass of NGC~7673 enclosed in the IUE slit is still M$_{total}^{ex}$ = 
1.1$_{-0.2}^{+1.5}\times$10$^{10}$ M$_{\odot}$. The best-fitting, multi-burst
SFH gives M$_{total}^{mb}$ = 3$_{-2}^{+4}\times$10$^{9}$ M$_{\odot}$,
a factor of 3 smaller than M$_{field}^{ex}$. This discrepancy is not 
surprising, since the star clusters are biased towards younger ages than
the exponentially decaying SFH and intrinsically brighter at ultraviolet
wavelengths; hence, the multi-burst SFH requires a lower mass to match
the observed FUSE/IUE spectrum. Clearly, the above values of M$_{total}^{ex}$ 
and M$_{total}^{mb}$ are a lower limit to the true stellar mass of NGC~7673, 
given that they have been measured for the area of the galaxy covered by the 
IUE slit and only from the UV portion of the galaxy spectrum. These 
mass estimates do not take into account the population of older/colder stars
emitting at optical and infrared wavelengths. 

For comparison, the HI mass
measured by Pisano et al. (2001) amounts to 4.1$\times$10$^{9}$ M$_{\odot}$ 
and the dynamical mass inferred from the HI line profile is 
2.5$\times$10$^{10}$ M$_{\odot}$. The latter has been estimated 
assuming an inclination angle $i$ = 45 $\pm$ 15 degrees and within 
a radius of 8.3 kpc, a factor of about 4 larger than the optical R$_{eff}$
of the galaxy (R$_{eff}$ = 1.9 kpc, Pisano et al. 2001). On the contrary,
the IUE slit covers only about half of the UV/optical extension of NGC~7673.
Therefore, scaling the stellar mass, either M$_{total}^{ex}$ or 
M$_{total}^{mb}$, by this factor of 2 would make it comparable with the
dynamical mass of the galaxy. 
\par\noindent
The exponentially decaying SFH with a present-day rate of $\sim$1 
M$_{\odot}$~yr$^{-1}$ as derived from the UV data would not be able to 
consume the HI gas in a Hubble time. If a rate of $\sim$5 M$_{\odot}$~yr$^{-1}$
is instead considered (as derived from the mid-IR and as an upper limit
to the rate estimated from the stellar populations),
the same SFH would exaust the HI gas still present in NGC~7673 in about
2 Gyr, possibly transforming this clumpy-irregular galaxy into an early-type
(this timescale does not take into account the gas returned by stars
via winds and supernova explosions).

\subsection{Clusters disruption}
According to Figure 10, a multi-burst SFH fits reasonably well the 
FUSE/IUE spectrum of NGC~7673. There are two ways of interpreting such a
star-formation history. A first approach considers the star clusters and 
the field population as formed through the same bursts of star 
formation. Alternatively,
the field population may have arised only from the disruption of the star
clusters.  How plausible this scenario is depends on the
dynamical evolution of star clusters. 

Two main timescales rule the dynamical state of a star cluster, namely the 
relaxation and the disruption time. The relaxation time is the time 
for a cluster to reach equilibrium between the kinetic energy distribution
of its stars and its potential well. Gravitational interactions among stars
of different mass have the effect of accelerating low-mass stars, which 
may leave the cluster (evaporation), while more massive stars sink into 
the inner region of the cluster leading to mass segregation and core-collapse.  
Wielen (1988) computed the relaxation time of an isolated cluster as a 
function of its mass $M_{cl}$, effective radius $R_{eff}$ and number of 
stars N$_{stars}$:  
\[
T_{rel} = \frac{3.6\times10^{-7}~N_{stars}}{(GM_{cl}/R_{eff}^3)^{0.5}} yr
\]
where G is the gravitational constant. According to this equation, low-mass 
clusters with a small radius have a short $T_{rel}$ (thus evaporate faster) 
than more massive and extended clusters. Larsen (2004) measured the size
of young star clusters in 18 nearby spiral galaxies, obtaining a mean 
$R_{eff} \simeq$ 3 pc, very weakly dependent on clusters age or mass.
Therefore, a star cluster with $M_{cl}$ = 10$^5$ M$_{\odot}$ and 
$R_{eff}$ = 3 pc has $T_{rel}$ = 1.6$\times$10$^6$~$N_{stars}$ yr, 
while $T_{rel}$ for a  cluster with $M_{cl}$ = 10$^7$ M$_{\odot}$ is
a factor of 10 shorter. The comparison of these timescales with the
cluster masses shown in Figure 6 clearly indicates that the star
clusters detected in NGC~7673 survive to their relaxation. A more
realistic computation of $T_{rel}$ has to take into account the gravitational
interaction between a star cluster and its parent galaxy, which truncates 
the cluster outskirts. Unfortunately, no tidal radius is known for young
star clusters, and in this case their $T_{rel}$ can not be estimated.

The disruption time is the time for tidal interactions with the parent
galaxy to destroy the star clusters, and depends on the clusters density, so that 
less-dense clusters
are more easily disrupted by their environment. Internal stellar evolution,
that can decrease the cluster density via stellar mass loss and supernova 
explosions, accelerates the cluster disruption. Boutloukos \& Lamers (2003)
determined the disruption time $T_{dis}$ of star clusters in four galaxies 
as a function of their initial mass function and their age distribution.
They found a large variation in $T_{dis}$ depending on the galaxy 
luminosity and morphology, from 40 Myr in M51 to 8 Gyr in the SMC. 
de Grijs et al. (2003) derived $T_{dis} \simeq$ 30 Myr for the star clusters 
in M82, with an uncertainty of a factor of 2. This value is comparable with
the age of 20 Myr computed for about half of the star clusters
in the area of NGC~7673 sampled by the IUE slit. This could then support the 
hypothesis whereby the field population older than 20 Myr formed in 
NGC~7673 through the disruption of star clusters. We could also speculate that 
the field population younger than 20 Myr is composed by star clusters too faint
to be detected in the HST images. Unfortunately, the sample of star
clusters is too small to allow for an accurate computation of $T_{dis}$ in
NGC~7673 with the method of Boutloukos \& Lamers. A similar scenario has been
also advocated by Chandar et al. (2005) to explain the nature of the diffuse 
ultraviolet light in 12 nearby starburst galaxies. 

\section{Conclusions}
We have used archival data publicly available for NGC~7673 to describe 
the recent star formation history that produced the young star clusters
and the field stellar population in this galaxy. We applied the photometric
dating technique of Pasquali et al. (2003) to the (F255W-F555W) and 
(F555W-F814W) colours of 50 star clusters identified with HST by Homeier et 
al. (2002), and made use of STARBURST99 models based on the Geneva 
evolutionary tracks with high stellar mass loss and computed with a 
LMC-like metallicity.
For the comparison between observed and synthetic colours we adopted three
extinction laws: Cardelli et al. (1989, for the Milky Way), Calzetti
(2001, for starburst galaxies) and Fitzpatrick (1986, for the
LMC), respectively. We have found that, independently of the assumed 
extinction law:
\par\noindent
- there are two prominent epochs of cluster formation, about 20 Myr and
2 Myr ago (see also Homeier et al. 2002), with somewhat minor events 
between 3 Myr and 6 Myr ago. 
\par\noindent
- The cluster intrinsic reddening E(B-V)$_i$ is lower than 0.4 mag; 
the large majority of clusters have 0 $\leq$ E(B-V)$_i \leq$ 0.25 mag.
\par\noindent
- The cluster mass is lower than 2$\times$10$^6$ M$_{\odot}$. 
\par
Out of these 50 star clusters, we have selected 31 placed within the boundaries
of the IUE large slit that was used to acquire the UV spectrum of NGC~7673
between 1150 \AA\/ and 3350 \AA. For each cluster, we built a synthetic
spectrum (with STARBURST99) corresponding to the age and mass derived from the
cluster photometry, and we estinguished it with each extinction law and its
corresponding  E(B-V)$_i$ as recovered from the colours measured for the cluster. 
All these 
31 synthetic spectra were then added together in the clusters integrated spectrum.
All the adopted extinction laws ensure a ratio of $\simeq$ 20$\%$ between 
the clusters integrated spectrum and the observed IUE spectrum at $\lambda$ = 
2578 \AA\/ (the mean wavelength of the F255W filter), in agreement with the
same ratio measured on the HST images by Homeier et al. ($\simeq$ 24$\%$). 
\par\noindent
The observed IUE spectrum (extended to 900 \AA\/ with the FUSE data collected
for NGC~7673) and the clusters integrated spectrum have allowed us to 
estimate the recent star-formation history of the field stellar population in
this galaxy. We have found that:
\par\noindent
- an exponentially decaying or a multi-burst star-formation history fits
equally well the observed FUSE/IUE spectrum.
\par\noindent
- The best-fitting exponentially decaying SFH is defined by an initial
star-formation rate $SFR_o \simeq$ 16 M$_{\odot}$~yr$^{-1}$ (Fitzpatrick)
or 13 M$_{\odot}$~yr$^{-1}$ (Calzetti), an e-folding
time $\tau \simeq$ 700 Myr (Fitzpatrick) or 900 (Calzetti), and started 
approximately 2 Gyr ago.
\par\noindent
- A multi-burst SFH assumes that the same bursts of star formation that 
produced the star clusters built up also the field population. The best-fitting
multi-burst SFH is characterised by a component as young as the star clusters
(hence younger than $\sim$40 Myr, see Figure 2) and a component as old
as 850 Myr (Fitzpatrick) or 450 Myr (Calzetti). The young component is a factor 
of about 2 - 3 brighter than the star clusters and the mass of the old component 
is estimated between 2 and 4 $\times$10$^9$ M$_{\odot}$
(i.e. about 100 - 200 times more massive than the star clusters together).
\par\noindent
The star-formation histories of the star clusters and the field population
together produce a number of ionizing photons large enough to sustain the
H$\alpha$ emission of NGC~7673, if not to allow also for an escape fraction
of $\sim$ 50$\%$.
The stellar mass of the galaxy enclosed in the IUE slit varies between
3$\times$10$^9$ M$_{\odot}$ (from the multi-burst SFH) to 1$\times$10$^{10}$ 
M$_{\odot}$ (from the exponentially decaying SFH). Given that the IUE slit
encompasses about half of the optical extension of the galaxy, a correction of
a factor of 2 should give roughly the total stellar mass of NGC~7673 locked 
up in stars younger than 2 Gyr.

\section*{Acknowledgments}
We would like to thank P. Benvenuti, J.S. Gallagher, F.C. van den Bosch
and H.-W. Rix for valuable discussions.

\clearpage

\setcounter{figure}{1}
\begin{figure*}
\centerline{\psfig{figure=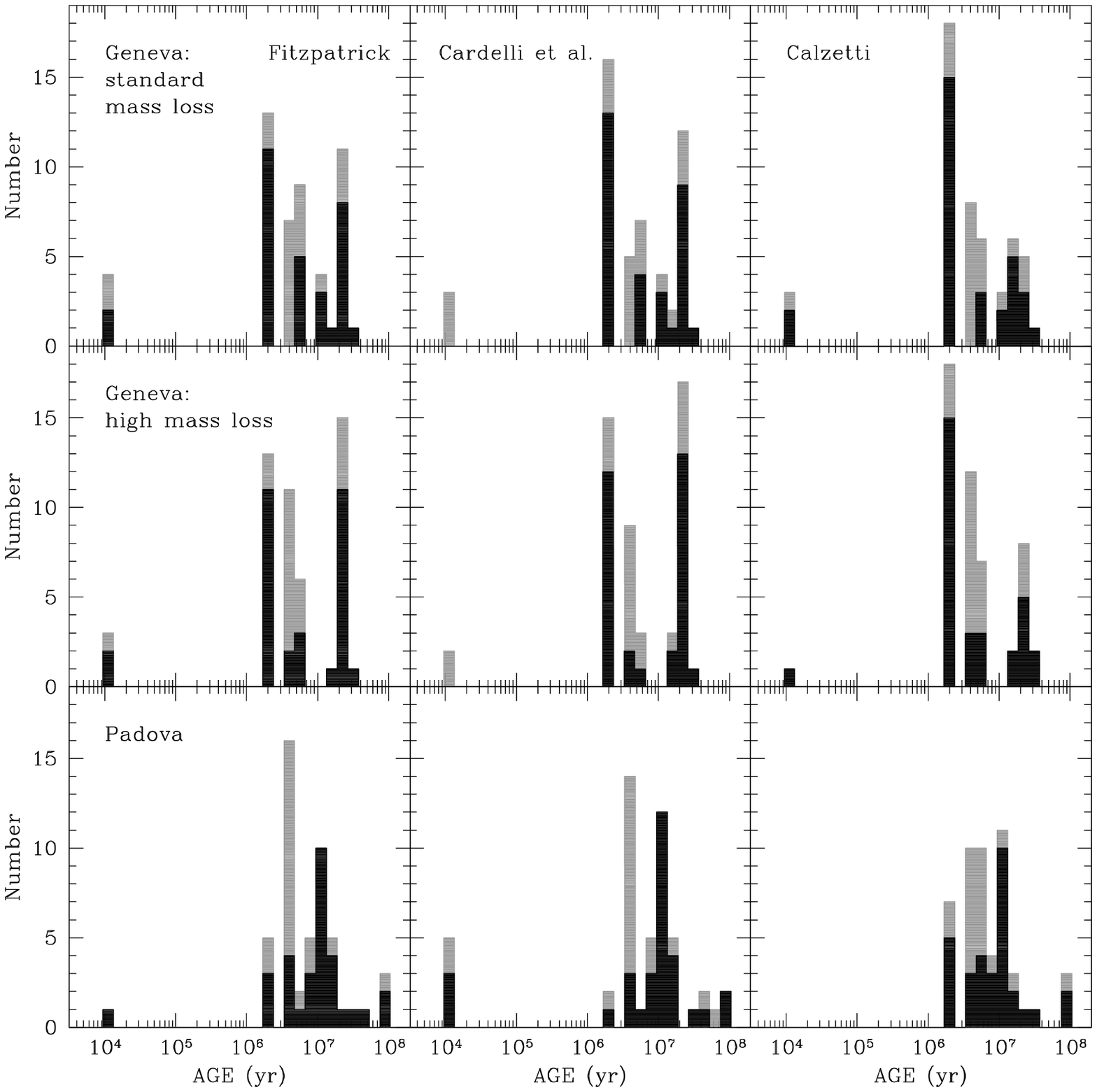,width=15cm}}
\caption{The clusters age distribution as a function of the adopted
evolutionary tracks and extinction laws. The grey histogram refers
to the full set of clusters (50) detected in F255W, F555W and F814W,
while the grey histogram represent only those (31) falling within
the IUE large aperture.}
\end{figure*}

\clearpage
\begin{figure*}
\centerline{\psfig{figure=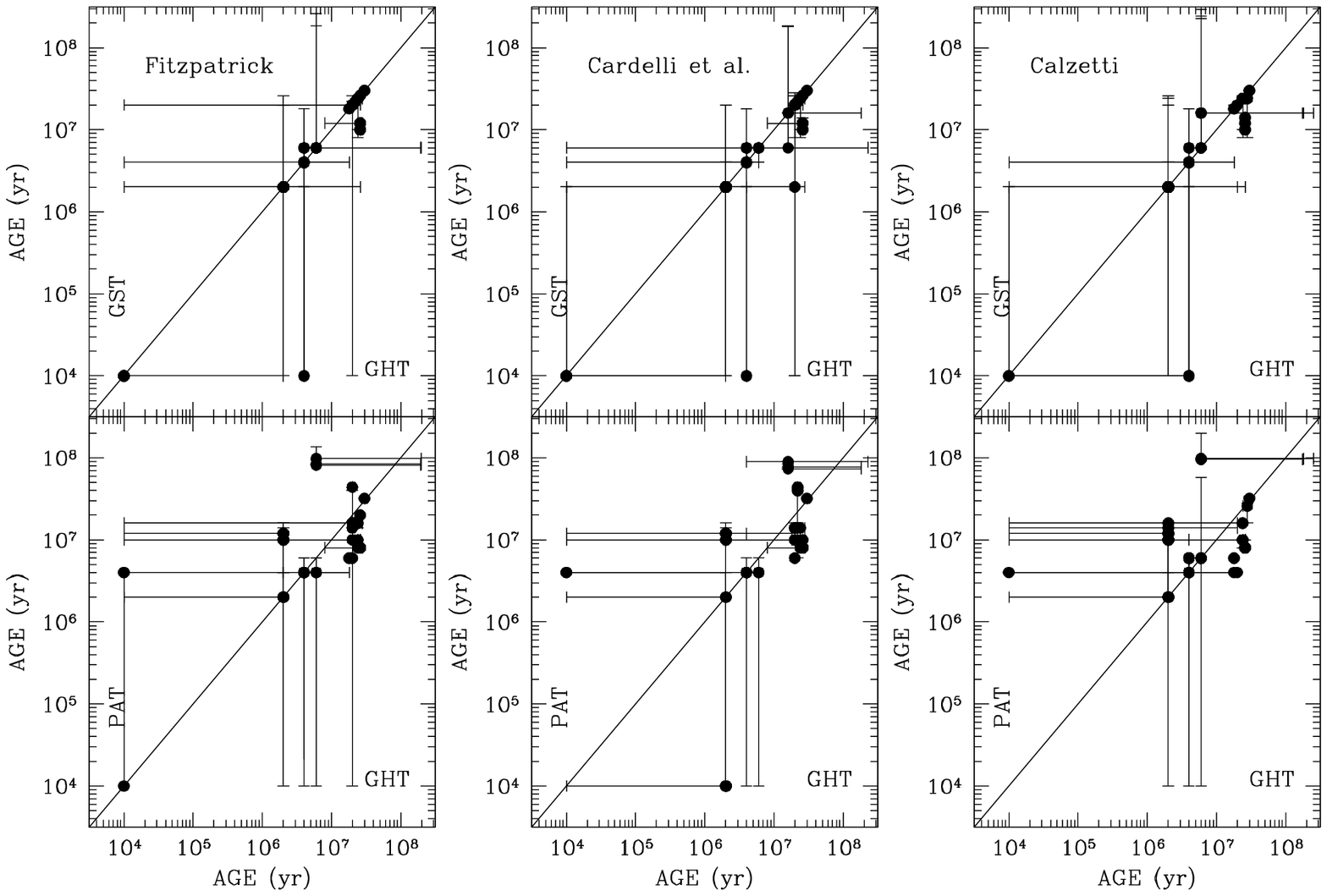,width=15cm}}
\caption{The clusters age obtained for different combinations of a set of 
evolutionary tracks with an extinction law are compared with each other,
and also with their 1$\sigma$ errors. The labels GST and GHT indicate the
Geneva Standard and High mass-loss tracks, respectively. PAT stands for the
Padova models including AGB evolutionary tracks (see Sect. 2.3).}
\end{figure*}

\clearpage
\begin{figure*}
\centerline{\psfig{figure=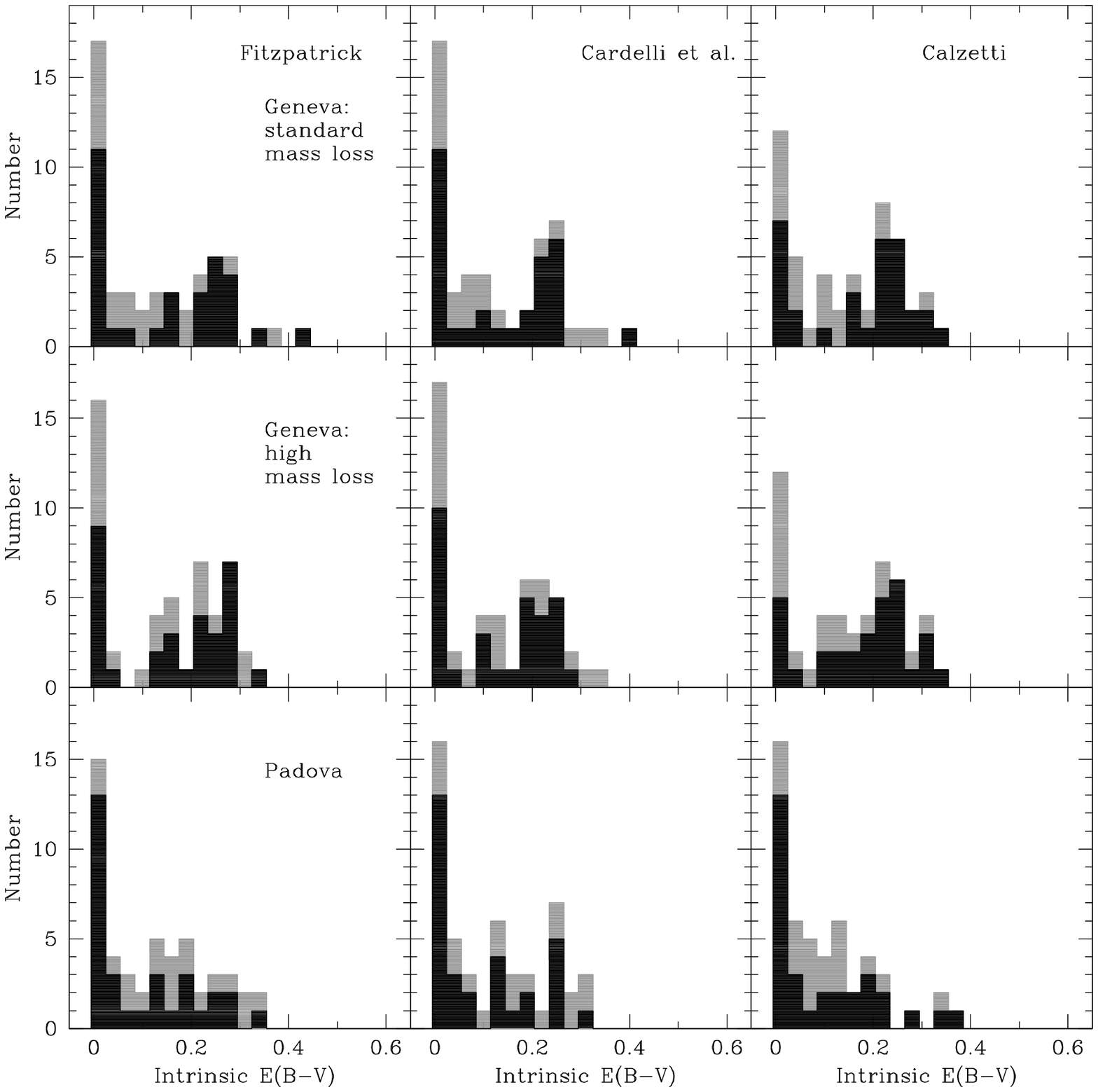,width=15cm}}
\caption{As in Figure 2, but for the clusters intrinsic reddening.}
\end{figure*}

\clearpage
\begin{figure*}
\centerline{\psfig{figure=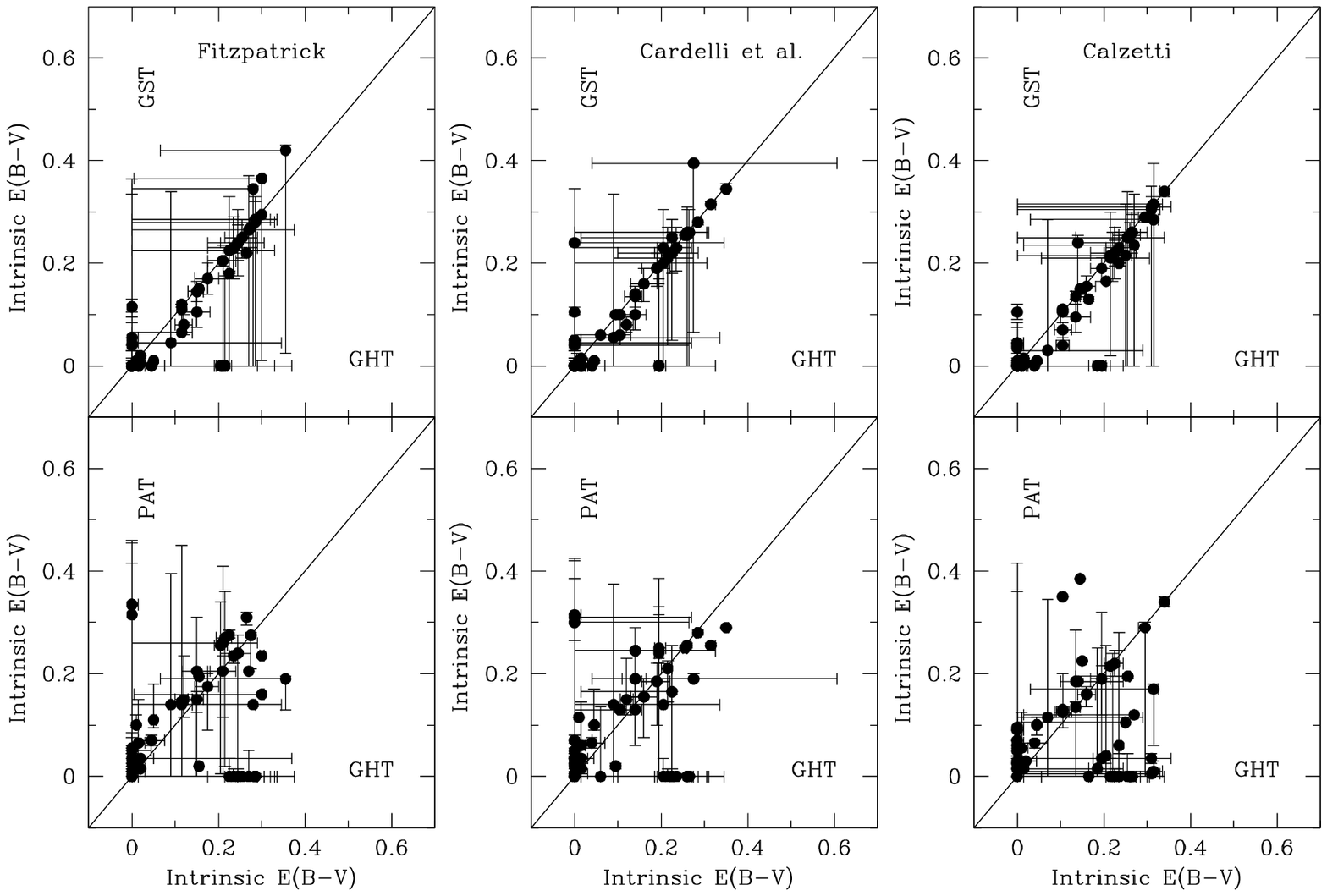,width=15cm}}
\caption{As in Figure 3, but for the clusters intrinsic reddening.}
\end{figure*}

\clearpage
\begin{figure*}
\centerline{\psfig{figure=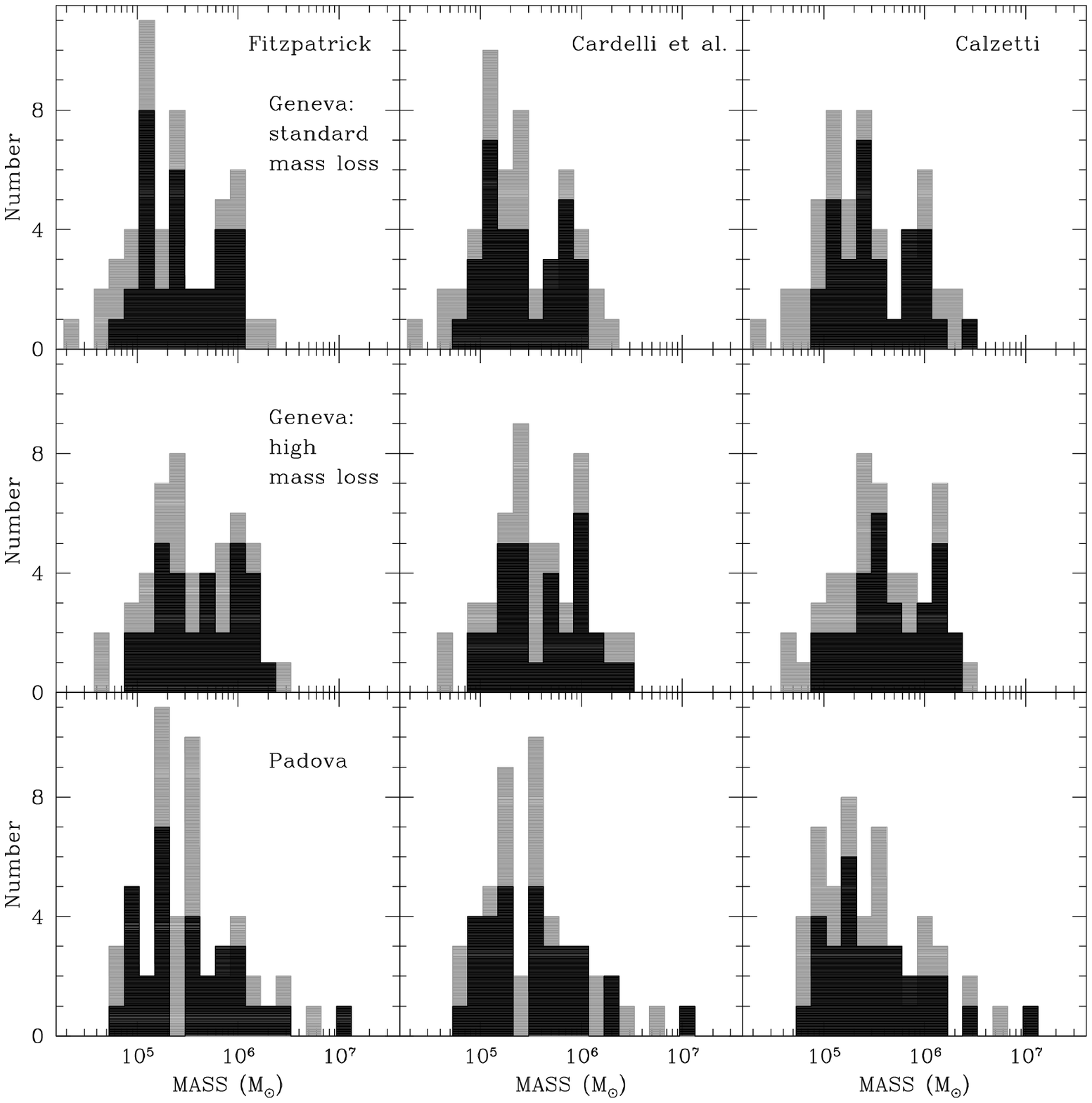,width=15cm}}
\caption{As in Figure 2, but for the clusters mass.}
\end{figure*}

\clearpage
\begin{figure*}
\centerline{\psfig{figure=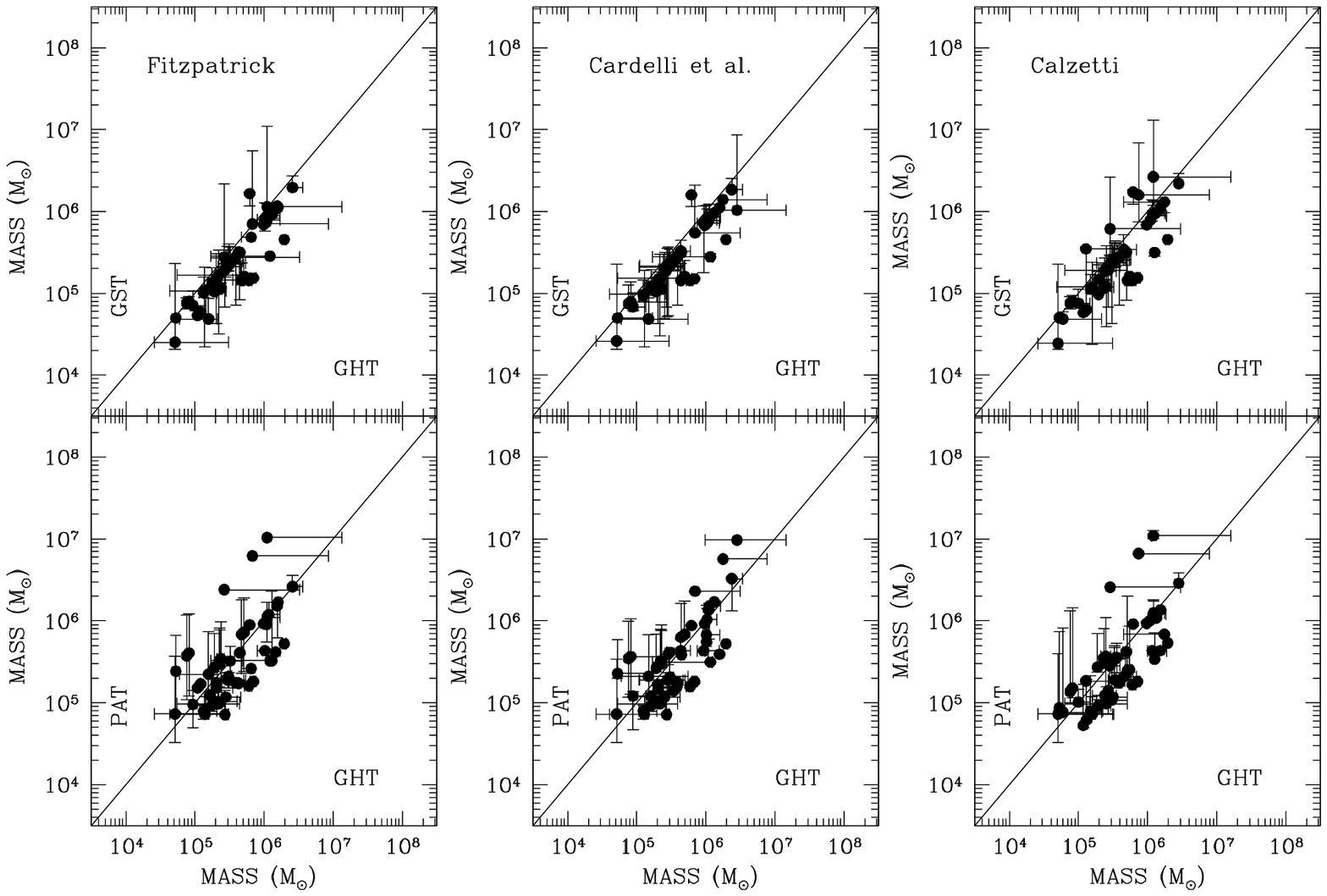,width=15cm}}
\caption{As in Figure 3, but for the clusters mass.}
\end{figure*}

\clearpage
\setcounter{figure}{8}
\begin{figure*}
\centerline{\psfig{figure=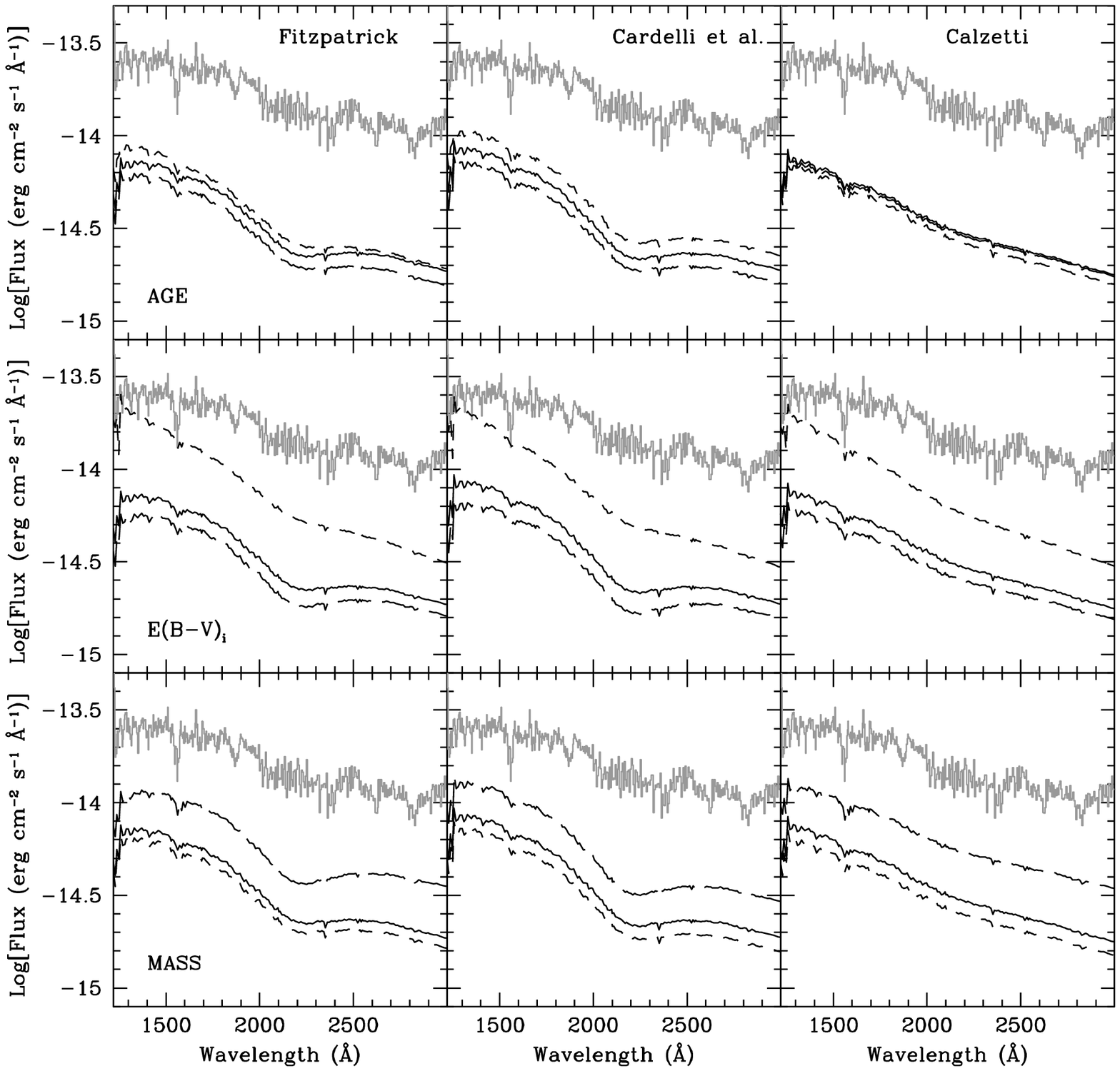,width=15cm}}
\caption{Comparison between the observed IUE spectrum of NGC~7673
(grey solid line) and the clusters integrated spectrum based on
the GHT (Geneva High mass-loss) tracks and reddened with three 
different extinction laws.
The black solid line represents the clusters integrated spectrum 
obtained for the best-fitting values of age, mass and E(B-V)$_i$, 
while the black short- and long-dashed lines trace, respectively, 
the clusters integrated spectrum derived for the lower and upper 
limits of these parameters.}
\end{figure*}

\end{document}